\documentclass[%
aps,
 pra,
 superscriptaddress,
 amsmath,amssymb,
reprint,%
]{revtex4-1}

\usepackage[utf8]{inputenc}
\usepackage{graphicx}
\usepackage{dcolumn}
\usepackage{bm}
\usepackage{slashbox}
\usepackage{braket}
\usepackage{bbold}
\usepackage{url}
\usepackage{verbatim}
\usepackage{lineno}
\usepackage{color}
\usepackage[normalem]{ulem}
\usepackage[colorinlistoftodos]{todonotes}

\newcommand{\kket}[1]{\ket{#1} \rangle}

\newcommand{\brakket}[1]{\braket{#1} \rangle}

\begin{document}



\title{Conditional probabilities of measurements, quantum time and the Wigner's friend case} 



\author{M. Trassinelli}
\email[]{martino.trassinelli@insp.jussieu.fr}
\affiliation {Institut des NanoSciences de Paris, CNRS, Sorbonne Université, Campus Pierre et Marie Curie, 75005 Paris, France}
\date{\today}

\begin{abstract}
Considering a minimal number of assumptions and in the context of the timeless formalism, 
conditional probabilities are derived for subsequent measurements in the non-relativistic regime.
Only unitary transformations are considered with detection processes described by generalized measurements (POVM).
One-time conditional probabilities are 
unambiguously
 derived via the Gleason-Bush theorem, including for puzzling cases like the Wigner's friend scenario where their form underlines the relativity aspect of measurements. 
No paradoxical situations emerge and the roles of Wigner and Wigner can be seen by his friend as being in a superposition.
\end{abstract}


\maketitle 

\section{Introduction} 

The measurement problem is one of the most fundamental issues in Quantum Mechanics on which the numerous interpretations and foundations differ. 
Von Neumann was the first to try to formalize the problem of measurement assuming that there are basically two types of evolution in quantum mechanics \cite{VonNeumann}. 
The first is unitary evolution, while the other is the collapse of the wave-function during the measurement from a superposition of states to only one of the studied systems.
Contrary to unitary evolution, the second measurement process, formalized by the \textit{projection postulate}, is irreversible, non-causal and, of course, non-unitary; it is generally viewed as the passage between the quantum and the classical realms.
When such a process is not postulated, different mechanisms can be evoked to justify it, like the coupling to the environment (decoherence) \cite{Zurek2003}, the addition of non-linearities in the evolution equation \cite{Ghirardi1979} and specific coupling to the gravitational fields \cite{Penrose1996}, among many other approaches.

If an ideal Ockham's razor would be applied to the different interpretations of Quantum Mechanics, only those with a minimal number of hypotheses should be considered (and that agree with experimental measurements of course).
Instead of postulating two type of natures, classical and quantum, the simplest approach is to consider only the second one; quantum phenomena cannot be explained by classical physics, whereas classical behavior can emerge from quantum systems. 
With the same minimalist approach, only one evolutionary process--the unitary processes--should be taken into account.
If only unitary transformations are considered, the measurement process has to be interpreted then as a unitary interaction between the detector system and the studied system. 
For this task, the standard approach is to describe the detection process by a positive valued operator (effect) in the context of the Positive Valued Operator Measure (POVM) framework \cite{AulettaQM,Busch}. 
In this context, non-ideal measurements can also be treated including destructive detections.

By considering the measuring apparatus as a quantum system, as well as the entire chain up to the observer's neurons, the measurement theory becomes part of the quantum theory of interacting compound systems.
In this framework, the correlations created by interactions between systems play a central role, like in Relational Quantum Mechanics  \cite{Rovelli1996,Rovelli2018}, in the original Everett formulation \cite{Everett1957} and its many-worlds theory derivation \cite{manyworlds}, in the Consistent Histories \cite{Griffiths,Omnes1992} , in the History Projection Operator formalism \cite{Isham1994} and in QBism \cite{Caves2002}.
This simplicity has however a price. 
The entanglement between observer and observable leads to different measurement results for independent, i.e. non-interacting, observers.
This relativity of what is normally considered universal remains a big conceptual leap for beings living in a world where quantum behaviors are barely visible. 
This discomfort manifests itself in paradoxes like in the case of the Wigner's friend scenario \cite{Wigner1995}, where two observers can have contradictory descriptions of the same system at the same moment.

Another thorny problem of the standard formulation of quantum mechanics is the special role occupied by the time coordinate $t$.
In the unitary evolution, $t$ is the evolutionary parameter; in the Schrödinger equation, time explicitly appears in the derivation with respect to it.
In both cases, time plays a special role in the description of a system's evolution.
Moreover, the formal description of a system before and after a measurement (via a short interaction) requires considering two distinct Hilbert spaces $\mathcal{H}_{in} \otimes \mathcal{H}_{out}$ \cite{Isham1994,Oreshkov2012,Flatt2017,Castro-Ruiz2018}.
This complexification and the central role of the time coordinate can be eliminated at once by including the measurement of time itself in a more general description of the studied interaction together with a clock system.
Originally designed to be compatible with general relativity, this formalism allows for considering dynamic processes without time as parameter.
This leads to a simpler formulation of the measurement process acting in a unique Hilbert space, where time is an observable among others resulting in a timeless description of dynamics.

Starting from a minimalistic description of nature using only quantum systems and unitary interactions, this article focuses on the formulation and properties of the conditional probability function in the quantum time formalism.
Since the early formulations from Page and Wootters in 1983 \cite{Page1983}, several works have discussed the conditional probability. 
This work takes advantage of the best of both the timeless and POVM formalisms to provide a coherent description of different measurements.

Differently from previous works \cite{Hellmann2007,Gambini2009,Giovannetti2015,Castro-Ruiz2020,Hohn2021,Baumann2021}, here the conditional probability for subsequent measurements is derived starting from first principles.  
More precisely, the most general measurement operator is built for the total Hilbert space that includes the clock system itself. 
By applying the Gleason-Bush theorem, the associate probability is derived.
As it will be show, in this way the probability expression is univocally defined.
In particular for Wigner's friend measurement scenarios, the associated probability expressions can be simply and unambiguously formulated.

In the following section, the basic notion of the quantum time/timeless formalism is introduced, including detector systems. 
In section \ref{sec:prob}, we derive the corresponding probability expressions and in Sec.~\ref{sec:wigner} we apply the developed tools to study the case of the Wigner's friend.
Section \ref{sec:conclusions} will be our conclusions.

\section{Non-relativistic quantum time (timeless evolution) formalism} 

Similarly to Refs.  \cite{Hellmann2007,Giovannetti2015,Baumann2021}, the Wheeler-DeWitt equation is taken as the starting point for describing the dynamics of the global state $\ket{\Psi}$ in the space-time continuum \cite{QG}: 
\begin{equation}
\hat H \kket{\Psi} = 0. \label{eq:WDW}
\end{equation}
$\hat H$ is the total Hamiltonian acting in the kinematic Hilbert space $\mathcal{K}$ where space-time coordinates are measured.
The notation with the double ket $\kket{\cdot}$ indicates the inclusion of a clock system in $\kket{\Psi}$ from which the time coordinate is measured.
In the following paragraphs, we implicitly assume $c = \hbar = 1$.
Solutions of the above equation are also solutions of 
\begin{equation}
 \kket{\Psi} = \int  d \alpha e^{- i \alpha \hat H} \kket{\Psi} = \mathbb{P} \kket{\Psi}. \label{eq:projector}
\end{equation} 
The operator  $\mathbb{P}$ can be considered as a ``projector''  from  $\mathcal{K}$ to the dynamical Hilbert space $\mathcal{H}$, the Hilbert space of the dynamical solutions \cite{QG}.

In the non-relativistic approximation where the perfect clock system $C$ is considered non-interacting with the system of interest $S$, the kinematic Hilbert space can be decomposed into $\mathcal{K} = \mathcal{H}_T \otimes \mathcal{H}_S$ where $\mathcal{H}_T$ and $\mathcal{H}_S$ are the Hilbert sub-spaces of the clock and the studied system, respectively.
The corresponding Hamiltonian is
\begin{equation}
\hat H = \hat p_T \otimes  \mathbb{1}_S +  \mathbb{1}_T \otimes \hat H_S \label{eq:NR_H}
\end{equation}
where $\hat H_T = \hat p_T$ consists of the conjugate operator of the time operator $\hat T$,  with $[\hat T, \hat p_T]=i$.
$\{ \ket{t}_T \}$ are the corresponding time states that form the base of $\mathcal{H}_T$, with $\hat T \ket{t}_T = t \ket{t}_T$ and $_T\!\braket{t | t'}_T = \delta (t -t')$.
A more general case where the clock can interact with the studied system or with another clock is treated in Refs.~\cite{Castro-Ruiz2017,Smith2019}.
The case of relative or non-ideal clocks is discussed in Refs. \cite{Loveridge2019,Rossi2020}. 

The total wave-function can be then decomposed into $\kket{\Psi} =  \ket{t}_T \otimes \ket{\psi(t)}_S$ where 
$\ket{\psi(t)}_S = _T\!\brakket{t | \Psi}$ 
is the wave-function obtained by the condition of measuring the time $t$ in the clock.
Consequently, we can also write
\begin{equation}
\kket{ \Psi} = \int dt \ket{t}_T \otimes \ket{\psi(t)}_S. \label{eq:t-decomp}
\end{equation}
With this notation, considering the term $_T\!\brakket{t| \hat H | \Psi}$ with Eqs.~\eqref{eq:WDW} and \eqref{eq:NR_H}, the 
standard form of the Schrödinger equation $i \frac{\partial}{\partial t} \ket{\psi(t)}_S = \hat H_S \ket{\psi(t)}_S$ is obtained. 
If $\hat H_S$ does not depend on the time operator $\hat T$, a solution of the equation is the unitary operator 
\begin{equation}
\hat U_S(t,t_0) = e^{-i \hat H_S (t - t_0)}. \label{eq:U}
\end{equation}

\section{Quantum measurement and probabilities} \label{sec:prob}

Similarly to the procedure described in Refs.~\cite{Hellmann2007,Giovannetti2015}, we consider here a measurement at a time $t_M$ consisting of a unitary interaction of a negligible duration between the system $S$ and a measuring system $M$.
The total Hamiltonian describing the system is
\begin{multline}
\hat H = \hat p_T \otimes  \mathbb{1}_S \otimes \mathbb{1}_M +  \mathbb{1}_T \otimes \hat H_S \otimes \mathbb{1}_M  \\
+ \hat V_{SM}\delta (\hat T-t_M) + \mathbb{1}_T \otimes \mathbb{1}_S \otimes \hat H_M. 
\label{eq:H}
\end{multline}
$V_{SM}$ represents the interaction between $\mathcal{H}_S$ and $\mathcal{H}_M$ at the time $t_M$, where $\hat H_M$ is the Hamiltonian of the detector itself.
With the detector space $\mathcal{H}_M$ we can associate the entire sub-system chain from the detector to the observer's brain.
The changes in $\mathcal{H}_S$ of a measurement with outcome $m$ is given by the positive-valued operator $\hat \Pi^m$ linked to the detector ancillary state $\ket{m}_M \in \mathcal{H}_M$.
The interaction between the detector and the system at time $t_M$ results in the unitary mapping
\begin{equation}
\ket{\psi(t_M)}_S \otimes \ket{r}_M \to \sum_m K_M^m \ket{\psi(t_M)}_S \otimes \ket{m}_M, \label{eq:map}
\end{equation}
where $\ket{r}_M$ is the ``ready'' detector state before the measurement, and $\hat K_M^m$ are the different Kraus operators corresponding to the outcomes $m$ with $\hat \Pi^m = (\hat K_M^m)^\dag \hat K_M^m$.
From Eq.~\eqref{eq:t-decomp}, the total state $\kket{\Psi}$ can be written as
\begin{multline}
\kket{\Psi} =  \int_{-\infty}^{t_M} dt \ket{t}_T \otimes \ket{\psi(t)}_S \otimes \ket{r}_M \\
+ \int_{t_M}^\infty dt \ket{t}_T \otimes \sum_i  \hat U_S(t,t_M) \hat K_M^i \ket{\psi(t_M)}_S \otimes \ket{i}_M, \label{eq:final_psi}
\end{multline}
where $\hat U_S$ is defined in Eq.~\eqref{eq:U}.

We consider now the case of two subsequent measurements: one measurement $M$ on $S$ with output $m$ operated at a time $t_M$ and another $N$ with output $n$ done at the time $t_N > t_M$ with no other interaction between the different sub-systems.
The corresponding Hamiltonian is 
\begin{multline}
\hat H = \hat p_T \otimes  \mathbb{1}_S \otimes \mathbb{1}_M \otimes \mathbb{1}_N +  \mathbb{1}_T \otimes \hat H_S \otimes \mathbb{1}_M \otimes \mathbb{1}_N \\
+ \hat V_{SM}\delta (\hat T-t_M) \otimes \mathbb{1}_N + \hat V_{SN}\delta (\hat T-t_N) \otimes \mathbb{1}_M \\
+ \mathbb{1}_T \otimes \mathbb{1}_S \otimes \hat H_M \otimes \mathbb{1}_N + \mathbb{1}_T \otimes \mathbb{1}_S \otimes \mathbb{1}_M \otimes \hat H_N
\end{multline}
and the associated wavefunction is \cite{Giovannetti2015,Castro-Ruiz2020},
\begin{multline}
\kket{\Psi} =  \int_{-\infty}^{t_M} dt \ket{t}_T \otimes \ket{\psi(t)}_S \otimes \ket{r}_M \otimes \ket{r}_N \\
+ \int_{t_M}^{t_N} dt \ket{t}_T \otimes \sum_i  \hat U_S(t,t_M) \hat K_M^i \ket{\psi(t_M)}_S \otimes \ket{i}_M \otimes \ket{r}_N \\ 
+  \int_{t_M}^{\infty} dt \ket{t}_T \otimes \sum_{i,j}  \hat U_S(t,t_N) \hat K_N^j \hat U_S(t_N,t_M) \hat K_M^i \ket{\psi(t_M)}_S \\
\otimes \ket{i}_M \otimes \ket{j}_N.\\ 
\label{eq:final_psi_2}
\end{multline}
Note that this timeless representation of the total wave-function by a series of superimpositions of system/detector entangled states is \textit{de facto} equivalent to the branched wavefunctions in Everett's original paper \cite{Everett1957}, where the central concepts turn around the relativity of states of measured and measuring systems.

How can we now associate a probability to the measurement outputs?
Different approaches have been developed in the past in the discussed context, as the definition via transition amplitudes in Quantum Gravity \cite{QG,Hellmann2007} or postulating the validity of the Bohr rule \cite{Giovannetti2015}.
Our approach consists in considering the general features of Hilbert spaces and requiring only basic properties of the probability function $\wp$.
For a Hilbert space with finite dimension $\mathcal{H}$, the Gleason-Busch theorem \cite{Gleason1957,Busch2003} demonstrates that the probability function is in fact univocally defined by the trace rule 
\begin{equation} 
\wp(a) = tr(\hat \rho\, \hat \Pi^a), \label{eq:trace}
\end{equation}
where $\hat \Pi^a$ is a positive-valued operator in $\mathcal{H}$ and $\hat \rho$ is the density matrix of a given state.
This is actually the same approach used by Pages and Wootters \cite{Page1983}, where however only projector operators are considered.
It is natural to extend this definition of probability to the infinite dimensional space $\mathcal{K}$.
Such an extension is in fact equivalent to the standard definition in Quantum Gravity.
In our case, a generic measurement with output $a$ must be represented by an operator $\hat \Pi_a$ in the kinematic Hilbert space $\mathcal{K} = \mathcal{H}_T \otimes \mathcal{H}_S \otimes \mathcal{H}_M \otimes \mathcal{H}_N$.  
Because there is only one clock subspace $\mathcal{H}_T$, within the minimal approach presented here, it is impossible to construct a two-time probability for the event ``$m$ at the time $t_1$  and $n$ at the time $t_2$''.
On the contrary, it is possible to consider the measurement ``$m \land  n \land t$'' $\equiv$  ``\textit{m AND n AND t}'' corresponding to the ``$m$ and $n$ results obtained at the time $t$'', when the two measurements occur at the times $t_M$ and $t_N$, respectively.
Like in Refs.~\cite{Giovannetti2015,Hellmann2007}, the formulation of such a probability implies that the two detector states $\ket{m}_M$ and $\ket{n}_N$ are stored in an internal memory that is read at the time $t> t_M, t_N$.
It is important to note that, in contrast to a previous work on subsequent measurements without timeless notation  \cite{Trassinelli2020b},
here the operator ``$\land$'' $\equiv$ ``AND'' is symmetric ($a \land b  = b \land a$). 
The measurement order is in fact defined by the Hamiltonian itself. 
We are not simply discussing $\wp(m \land n \land t)$, but we are implicitly considering the probability $\wp(m \land n \land t | M_{t_M} \land N_{t_N})$, where we 
explicitly indicate the dependence on
 the prior information of the Hamiltonian structure, with the measurements $M$ at the time $t_M$ and $N$ at $t_N$. 
More complex scenarios with an undefined measurement order in the Hamiltonian can be also considered \cite{Castro-Ruiz2020,Paunkovic2020} but they are not discussed here.

With respect to two-time probabilities, a naive form, based in wavefunction two-time collapse, does not reproduce the correct propagator \cite{Kuchar2011}. 
An alternative solution to this problem has been provided by Dolby \cite{Dolby2004}. 
However, it has been demonstrated \cite{Hellmann2007} that this solution is in conflict with predictions of standard quantum mechanics.
Recently, a new proposition of two-time probability that resolves past criticisms has been presented by Höhn and collaborators \cite{Hohn2021}.
The operators corresponding to the two measurements at the two times are built without violating Eqs.~\eqref{eq:WDW} and \eqref{eq:projector}, i.e. keeping the system in the physical Hilbert space $\mathcal{H}$. 
In the present work, we take a simpler approach considering measurement memories, such that the time coordinate is read only once. 
In this way, we can easily define probability outputs from a generic one-time operator $\hat \Pi_a  = \hat \Pi^{m,n,t}_{MN}  \in \mathcal{K}$ instead of $\mathcal{H}$.
Even if the total wave-function is not part of  $\mathcal{H}$ after the measurement ($\kket{\Psi} \to \hat K_a \kket{\Psi} \notin \mathcal{H}$, where $\hat K_a$ is the Kraus operator associated to $\hat \Pi_a = \hat K_a^{} \hat K_a^\dag$), the probability $\wp(a)$ is well defined by the Gleason-Bush theorem (Eq.~\eqref{eq:trace}).

It should be emphasized that there is no direct measurement on $S$ but only via the ancillary detector states.
The operator associated with this measurement is 
$\hat \Pi^{m,n,t}_{MN} = \ket{t}_T {^{}_T}\! \bra{t}  \otimes \mathbb{1}_S \otimes \ket{m}_M {^{}_M}\! \bra{m} \otimes \ket{n}_N {^{}_N}\! \bra{n} $ 
that leads to
\begin{multline}
\wp(m \land n \land t) = tr(\hat \rho\, \hat \Pi^{m,n,t}_{MN})  =  ||  ( _T \bra{t} \otimes _M\! \bra{m}  \otimes _N\! \bra{n}) \kket{\Psi} || ^2  \\
=
\begin{cases}
|| ^{}_M\! \braket{m | r}_M\ ^{}_N\! \braket{n | r}_N \ket{\psi (t_M)}_S||^2 & \text{ if } t < t_M,\\
|| ^{}_N\! \braket{n | r}_N  \hat K^m_M \ket{\psi (t_M)}_S||^2 & \text{ if } t_M \le t < t_N, \\
|| \hat K^n_N \hat U_S(t_N,t_M)  \hat K^m_M \ket{\psi (t_M)}_S||^2 & \text{ if } t \ge t_N.\\
\end{cases}
\end{multline}

With two separate detections, we can now discuss the conditional probabilities of the different outcomes.
The conditional probabilities and $\wp(n | m \land t)$ can be calculated with the standard procedure from the Bayes’ rule  and the probabilities $\wp(n \land t)$ and $\wp(m \land t)$ by applying Eq.~\eqref{eq:trace} with the corresponding operators.
We obtain in particular
\begin{equation}
\wp(n | m \land t) = \frac{\wp(m \land n \land t)}{\wp( m \land t)} = \frac{ tr(\hat \rho\, \hat \Pi^{m,n,t}_{MN})} {tr(\hat \rho\, \hat \Pi^{m,t}_{M})}
= \wp(n \land t| m \land t),
\label{eq:conditional}
\end{equation}
where $\Pi^{m,t}_M = \ket{t}_T {^{}_T}\! \bra{t}  \otimes \mathbb{1}_S \otimes \ket{m}_M {^{}_M}\! \bra{m} \otimes \mathbb{1}_N.$
For $\wp(m | n \land t)$ a similar equation can be derived.
The validity of the last equivalence in the above formula is justified by the presence of the time measurement operator in both numerator and denominator.
Note that $\wp(n \land t | m ), \wp(m \land t | n )$ probabilities could also be used to build conditional forms.
In this case, their evaluation is more complex due to the appearance of $\wp(n)$ and $\wp(m)$ in the denominators.
It requires in fact the marginalization with respect the time operator basis $\{ \ket{t}_T \}$ and consequently its normalization.
Such a normalization is discussed  Ref.~\cite{Giovannetti2015} but is not considered here. When only $\wp(m \land n \land t)$ and $\wp(n \land t)$ are used to derive $\wp(m | n \land t)$, in fact no problems of normalization are encountered. 

\section{Wigner's friend scenario and conditional probability} \label{sec:wigner} 

\subsection{Formulae} 

With the notation and formulas presented above, we can now consider the case of Wigner's friend.
This measurement scenario, introduced for the first time by Wigner in 1961 \cite{Wigner1995}, is constituted by an observer $W$ (Wigner) that observes another observer $F$ (his friend) who performs a quantum measurement on a physical system $S$.
The friend makes a measurement of the system and records a well-defined result. 
During this measurement, and just after, the system/friend ensemble is isolated from Wigner.
For $W$, the measurement of $F$ on $S$ becomes entangled and both are described by a superposition of the different possible final states.
Paradoxically, $W$ and $F$ have at the same time a different description of $S$.
In particular, when quantum and classical worlds are both considered at the same time, with the collapse of the wavefunction delimiting them, the collapse itself is not well defined (especially when consciousness is considered as the collapse trigger like in Wigner's original argument).
The Wigner's friend scenario with the timeless formalism has been extensively discussed in a recent work of Baumann, Brukner and collaborators \cite{Baumann2021}, where several forms of probabilities are considered.
The notation presented here leads, in contrast, to a unique form of the probability.
Its derivation is presented in detail in the following paragraphs applying the results presented above.

As in Ref.~\cite{Baumann2021}, without losing in generality and to keep simple expressions, we consider no-free dynamics for three systems: the studied system $S$, Wigner's friend $F$, and Wigner $W$.
The corresponding Hamiltonian, relative to a measurement $M$ at the time $t_M$ between the system and the friend, and a measurement $N$ at the time $t_N$ between Wigner and the ensemble friend-system, is 
\begin{multline}
H = \hat p_T \otimes  \mathbb{1}_S \otimes \mathbb{1}_F \otimes \mathbb{1}_W +  \mathbb{1}_T \otimes \hat H_S \otimes \mathbb{1}_F \otimes \mathbb{1}_W  \\
+ \hat V_{SF}\delta (\hat T-t_M) \otimes \mathbb{1}_W + \hat V_{SFW}\delta (\hat T-t_M)\\
+ \mathbb{1}_T \otimes \mathbb{1}_S \otimes \hat H_F \otimes \mathbb{1}_W + \mathbb{1}_T \otimes \mathbb{1}_S \otimes \mathbb{1}_F \otimes \hat H_W.
\end{multline}
We consider a system initial condition described by 
\begin{equation} \label{eq:psi_S}
\ket{\psi}_S = a \ket{\uparrow}_S + b \ket{\downarrow}_S
\end{equation}
with $|a|^2 + |b|^2 = 1$.  
For Wigner, after the friend's measurement, $S$ and $F$ are described by
\begin{equation}
\ket{\psi}_S \otimes \ket{\varphi}_F = a \ket{\uparrow}_S \otimes \ket{\uparrow}_F + b \ket{\downarrow}_S \otimes \ket{\downarrow}_F.
\label{eq:ent_SF}
\end{equation}
We consider also that Wigner's measurement on $F$ and $S$ consists in the detection of the ancillary state $\ket{yes}_W$, which corresponds to the operator on the system-friend space $\hat \Pi_N^{yes} = \ket{yes}_{SF}\, ^{}_{SF}\bra{yes} \in \mathcal{H}_S \otimes \mathcal{H}_F$ with
\begin{equation}
\ket{yes}_{SF} = \alpha \ket{\uparrow}_S \otimes \ket{\uparrow}_F  + \beta \ket{\downarrow}_S \otimes \ket{\downarrow}_F, 
\label{eq:yes}
\end{equation}
with $|\alpha|^2 + |\beta|^2 = 1$, and its complementary operator $\hat \Pi_N^{no} = \mathbb{1}_S \otimes  \mathbb{1}_F -\hat \Pi_N^{yes}$.

The global wavefunction for these two measurements is similar to that of Eq.~\eqref{eq:final_psi_2}:
\begin{multline}
\kket{\Psi} =  \int_{-\infty}^{t_M} dt \ket{t}_T \otimes \ket{\psi}_S \otimes \ket{r}_F \otimes \ket{r}_W  \\
+ \int_{t_M}^{t_N} dt \ket{t}_T \otimes \sum_f  \hat K_M^f \ket{\psi}_S \otimes \ket{f}_F \otimes \ket{r}_W \\
+ \int_{t_N}^{\infty} dt \ket{t}_T \otimes \sum_{f,w}  \hat K_N^w \hat K_M^f \ket{\psi}_S \otimes \ket{f}_F \otimes \ket{w}_W.
\label{eq:final_psi_wigner}
\end{multline}
The probability of having the measurement results $w$ and $f$ at the time $t$ is associated to the operators $
\hat \Pi^{f,w,t}_{MN} = \ket{t}_T {^{}_T}\! \bra{t}  \otimes \mathbb{1}_S \otimes \ket{f}_F {^{}_F}\! \bra{f} \otimes \ket{w}_W {^{}_W}\! \bra{w}$.
Its values are 
\begin{multline}
\wp(f \land w \land t) =  tr(\hat \rho\, \hat \Pi^{f,w,t}_{MN})  =
|| ( _T \bra{t} \otimes _F\! \bra{f}  \otimes _W\! \bra{w}) \kket{\Psi} ||^2  = \\
\begin{cases}
|| ^{}_F\! \braket{f | r}_F\ ^{}_W\! \braket{w | r}_W \ket{\psi}_S||^2 & \text{ if } t < t_M,\\
|| ^{}_W\! \braket{w | r}_W  \hat K^f_M \ket{\psi}_S||^2 & \text{ if } t_M \le t < t_N, \\
||^{}_F\! \bra{f} \hat K^w_N  \sum_{f'} \hat K^{f'}_M \ket{\psi}_S \otimes \ket{f'}_F||^2 & \text{ if } t \ge t_N,
\end{cases}
\end{multline}
where $\ket{r}_F$ and $\ket{r}_W$ are the ``ready'' state of the friend and of Wigner. 
The associated Kraus operators are 
\begin{equation} \label{eq:Ks}
  \begin{split}
  &\hat K_M^{\uparrow} = \ket{\uparrow}_S {}^{}_S\! \bra{\uparrow},  \\
  &\hat K_M^{\downarrow} = \ket{\downarrow}_S {}^{}_S\! \bra{\downarrow}, \\
  &\hat K_N^{yes} = \ket{yes}_S \ket{yes}_F {}^{}_S\!\bra{yes} {}^{}_F\!\bra{yes}\\
  &= ( \alpha \ket{\uparrow}_S  \ket{\uparrow}_F + \beta \ket{\downarrow}_S \ket{\downarrow}_F ) (\alpha^* {}^{}_S\! \bra{\uparrow}  {}^{}_F\! \bra{\uparrow} + \beta^*\ {}^{}_S\! \bra{\downarrow} {}^{}_F\! \bra{\downarrow}), \\
  &\hat K_N^{no} = \ket{no}_S {}^{}_S\!\bra{no} \\
  &= ( -\beta^* \ket{\uparrow}_S  \ket{\uparrow}_F + \alpha^* \ket{\downarrow}_S ket{\downarrow}_F) (-\beta {}^{}_S\! \bra{\uparrow}  {}^{}_F\! \bra{\uparrow}  + \alpha {}^{}_S\! \bra{\downarrow} {}^{}_F\! \bra{\downarrow}),
   \end{split}
\end{equation}
where we omitted the tensor product symbol to keep the formulas compact.
Differently from the general case of subsequent measurements treated in Sec.~\ref{sec:prob}, where all $\hat K$ operators were acting only on $\mathcal{H}_S$ (like $\hat K_{M}^f$ as well), $\hat K_{N}^w$ is operating on  $\mathcal{H}_S \otimes \mathcal{H}_F$.
This implies the more complex expression of $\wp(f \land w \land t)$ where the tensor product between $\ket{\psi}_S$ and $\ket{f'}_F$ is still present because, differently from $\ket{w}_W$ basis, the orthogonality on the friend basis ${}_F\!\braket{f'|f}_F = \delta_{f f'}$ cannot be directly used.
More explicitly, for $t \ge t_N$ we have
\begin{multline}
\wp(f \land w \land t) =  ||^{}_F\! \bra{f} \hat K^w_N  \sum_{f'} \hat K^{f'}_M \ket{\psi}_S \otimes \ket{f'}_F||^2 \\
= ||^{}_F\! \bra{f} \hat K^w_N  ( a \ket{\uparrow}_S \otimes \ket{\uparrow}_F  + b \ket{\downarrow}_S \otimes \ket{\downarrow}_F) ||^2. \label{eq:pwft_explicit}
\end{multline}

Considering the above equation and the explicit form of $\hat K^w_N$ (Eqs.~\eqref{eq:Ks}), 
the different possible values of $\wp(f \land w \land t)$ 
are obtained and
summarized in Tab.~\ref{tab:Ptot}.

\begin{table}
\caption{\label{tab:Ptot} $\wp(f \land w \land t)$ values for $t>t_N$.}
\begin{tabular}{c|cc}
\hline
\hline
 \backslashbox{f}{w} & yes &  no\\
 \hline
$\uparrow$ & $  | \alpha | ^2 | a \alpha^* + b \beta^* |^2$  & $| \beta | ^2 | a \beta - b \alpha |^2$ \\
$\downarrow$ & $ | \beta | ^2 | a \alpha^* + b \beta^* |^2$ & $ | \alpha | ^2 | a \beta - b \alpha |^2$ \\
\hline
\hline
\end{tabular}
\end{table}

Similarly to Eq.~\eqref{eq:conditional}, the conditional probabilities $\wp(w | f \land t)$ and $\wp(f | w \land t)$ can be calculated from 
\begin{equation}
 \wp(w | f \land t) =
 \frac{\wp(f \land w \land t)}{\wp(f  \land t)} = \frac{tr (\hat \rho\, \hat \Pi^{f,w,t}_{MN}) }{tr (\hat \rho\, \hat \Pi^{f,t}_{M})} 
\end{equation}
and
\begin{equation}
\wp(f | w \land t) = 
 \frac{\wp(f \land w \land t)}{\wp(w  \land t)} = \frac{tr (\hat \rho\, \hat \Pi^{f,w,t}_{MN}) }{tr (\hat \rho\, \hat \Pi^{w,t}_{N})} 
\end{equation}
where $\hat \Pi^{f,t}_{M} = \ket{t}_T {^{}_T}\! \bra{t}  \otimes \mathbb{1}_S \otimes \ket{f}_F {^{}_F}\! \bra{f} \otimes \mathbb{1}_W $ and $ \hat \Pi^{w,t}_{N} = \ket{t}_T {^{}_T}\! \bra{t}  \otimes \mathbb{1}_S \otimes \mathbb{1}_F \otimes \ket{w}_W {^{}_W}\! \bra{w}$.
To deduce their values, $\wp(w  \land t)$ and $\wp(f  \land t)$ have to be evaluated first.
For $\wp(w  \land t)$ we have 
\begin{multline} \label{eq:pwt}
\wp(w \land t) =  tr(\hat \rho\, \hat \Pi^{w,t}_{N})  =
|| ( _T \bra{t}  \otimes _W\! \bra{w}) \ket{\Psi} ||^2  = \\
\begin{cases}
|| ^{}_W\! \braket{w | r}_W \ket{\psi}_S \otimes \ket{r}_F ||^2 & \text{ if } t < t_M,\\
|| ^{}_W\! \braket{w | r}_W  \sum_f \hat K^f_M \ket{\psi}_S \otimes \ket{f}_F ||^2 & \text{ if } t_M \le t < t_N, \\
|| \hat K^w_N  \sum_f \hat K^f_M \ket{\psi}_S \otimes \ket{f}_F  ||^2 & \text{ if } t \ge t_N, 
\end{cases}
\end{multline}
%
where we explicitly used the expression of $\kket{\Psi}$ from Eq.~\eqref{eq:final_psi_wigner}.
For the case $t>t_N$, we obtain an expression very similar to Eq.~\eqref{eq:pwft_explicit}
\begin{equation}
\wp( w \land t) =   ||\hat K^w_N  ( a \ket{\uparrow}_S \otimes \ket{\uparrow}_F  + b \ket{\downarrow}_S \otimes \ket{\downarrow}_F) ||^2. \label{eq:pwt_explicit}
\end{equation}
where, differently from $\wp(f \land w \land t)$, the projection to a particular $\ket{f}_F$ is absent.
The different values of $\wp( w \land t)$ are reported in Tab.~\ref{tab:Pw_partial}.

\begin{table}
\caption{\label{tab:Pw_partial} Values of $\wp(w \land t)$  for $t>t_N$.}
\begin{tabular}{c|c}
\hline
\hline
w & $\wp(w \land t)$  \\
\hline
yes & $ | a \alpha^* + b \beta^* |^2$ \\
 no & $  | a \beta - b \alpha |^2 $ \\
\hline
\hline
\end{tabular}
\end{table}

\begin{table}
\caption{\label{tab:Pf_partial} Values of  $\wp(f \land t)$ for $t>t_N$.}
\begin{tabular}{c|c}
\hline
\hline
f &  $\wp(f \land t)$ \\
\hline
 $\uparrow$ & $ | \alpha | ^2 | a \alpha^* + b \beta^* |^2 + | \beta | ^2 | a \beta - b \alpha |^2 $ \\
 $\downarrow$ & $ | \beta | ^2 | a \alpha^* + b \beta^* |^2 + | \alpha | ^2 | a \beta - b \alpha |^2$ \\
\hline
\hline
\end{tabular}
\end{table}

For $\wp(f  \land t)$ we have
\begin{multline} \label{eq:pft}
\wp(f \land t) =  tr(\hat \rho\, \hat \Pi^{w,t}_{M})  =
|| ( _T \bra{t} \otimes _F\! \bra{f} ) \kket{\Psi} ||^2  = \\
\begin{cases}
|| ^{}_F\! \braket{f | r}_F \ket{\psi}_S||^2 \otimes \ket{r}_W &\text{ if } t < t_M,\\
|| \hat K^f_M \ket{\psi}_S \otimes \ket{r}_W ||^2  &\text{ if } t_M \le t < t_N, \\
||\sum_w {}^{}_F\! \bra{f}  \hat K^w_N  \sum_{f'} \hat K^{f'}_M \ket{\psi}_S &\hspace{-0.25cm} \otimes \ket{f'}_F \otimes  \ket{w}_W||^2
\end{cases}\\
 \text{ if }  t \ge  t_N. \hspace{0.55cm}
\end{multline}
%
Using the properties of the tensor product and orthogonality of the Wigner's basis $_W\!\braket{w | w'}_W = \delta_{w w'}$, we can sort the sum over $w$ and obtain
\begin{equation}
\wp(f  \land t)  = \sum_w ||^{}_F\! \bra{f}  \hat K^w_N  \sum_{f'} \hat K^{f'}_M \ket{\psi}_S \otimes \ket{f'}_F \otimes  \ket{w}_W||^2. \label{eq:pft_explicit}
\end{equation}
The different values of $\wp( f \land t)$
can be obtained from the sum over $w$ of single values of $\wp( f \land w \land t)$ (Tab.~\ref{tab:Ptot}) and are reported in Tab.~\ref{tab:Pf_partial}.

The corresponding conditional probabilities $\wp(w | f \land t)$ and $\wp(f | w \land t)$  can now be calculated and are reported in Tabs.~\ref{tab:Pw_cond} and \ref{tab:Pf_cond}, respectively.
 
\begin{table}
\caption{\label{tab:Pw_cond} Values of $\wp(w | f \land t)$ for $t>t_N$.}
\begin{tabular}{c|cc}
\hline
\hline
 \backslashbox{f}{w} & yes &  no\\
 \hline
$\uparrow$ & $\frac{| \alpha | ^2 | a \alpha^* + b \beta^* |^2}{| \alpha | ^2 | a \alpha^* + b \beta^* |^2 + | \beta | ^2 | a \beta - b \alpha |^2 }$  & 
$\frac{| \beta | ^2 | a \beta - b \alpha |^2}{| \alpha | ^2 | a \alpha^* + b \beta^* |^2 + | \beta | ^2 | a \beta - b \alpha |^2 }$ \\
$\downarrow$ & $\frac{| \beta | ^2 | a \alpha^* + b \beta^* |^2}{ | \beta | ^2 | a \alpha^* + b \beta^* |^2 + | \alpha | ^2 | a \beta - b \alpha |^2}$ &
 $\frac{| \alpha | ^2 | a \beta - b \alpha |^2}{ | \beta | ^2 | a \alpha^* + b \beta^* |^2 + | \alpha | ^2 | a \beta - b \alpha |^2}$ \\
\hline
\hline
\end{tabular}
\end{table}

\begin{table}
\caption{\label{tab:Pf_cond} Values of $\wp(f | w \land t)$ for $t>t_N$.}
\begin{tabular}{c|cc}
\hline
\hline
 \backslashbox{f}{w} & yes &  no\\
 \hline
$\uparrow$ & $ | \alpha |^2$  & $ | \beta |^2$ \\
$\downarrow$ & $| \beta |^2$ & $ | \alpha |^2$ \\
\hline
\hline
\end{tabular}
\end{table}

%


\subsection{Discussion} 

As can be seen
from above equations and the corresponding values in the different tables,
the probability function is defined without ambiguities 
for all cases
and is automatically normalized.
In particular
$\sum_w \wp(w | f \land t) = 1$ and $\sum_f \wp(f | w \land t) = 1$.
For defined values of $f$ and $w$, the value of $\wp(f \land w \land t)$  is simply defined by the the initial state (Eq.~\eqref{eq:psi_S}) and the states measured by Wigner (Eq.~\eqref{eq:yes}).
For $\wp(w  \land t)$ and $\wp(f  \land t)$ the situation is more complex and interesting. 
From the Wigner point of view, 
the friend and the system are described by a superimposition of states.  
The superimposition condition is
represented by the sum of terms $\hat K^f_M \ket{\psi}_S \otimes \ket{f}_F$ in Eq.~\eqref{eq:pwt}.
Similarly, the friend sees the state of Wigner as a superposition represented by the sum of the terms $\hat K^f_M \ket{\psi}_S \otimes \ket{f}_F$ in Eq.~\eqref{eq:pft}.
Because of the presence of the tensor product involving $\ket{w}_W$ basis, for $\wp(f \land | t) $ the square modulus of the sum of states in Eqs.~\eqref{eq:pft}, can be written as a sum of square moduli (Eq.~\eqref{eq:pft_explicit}).
This is not true for $\wp(w \land | t) $ because of the more complex expression due to the operators $\hat K^w_N$ acting on the friend Hilbert space $\mathcal{H}_F$ in addition to the system space $\mathcal{H}_S$.
The final expressions of $\wp(w | f \land t)$ and $\sum_f \wp(f | w \land t)$  are equivalent to the ones presented in Baumann et al. \cite{Baumann2021} for a specific choice of the probability definition (definition 2b).
Unlike the work presented in Ref.~ \cite{Baumann2021} where different timeless probability definitions are discussed for the Wigner’s friend case, a unique definition of probability emerges here, where all systems are considered to be quantum in nature (without any classical realm) and interacting with each other by unitary operators.

The presence of different superposition combinations for the probabilities $\wp(w \land t)$ or $\wp(f \land t)$ (Eqs.~\eqref{eq:pft},\eqref{eq:pwt}) is a manifestation of the different descriptions of the same system $S$ from Wigner and the friend.
The probabilities depend 
in fact
only on the 
mutual
relations $S-W$, $S-F$ and $F-W$.
In the last years, no-go theorems evoking situations similar to the Wigner's friend scenario have been formulated \cite{Frauchiger2018,Brukner2018,Bong2020} and experimentally tested \cite{Proietti2019,Bong2020}.
They demonstrated that one of the following assumptions has to be violated:  i) the universal validity of quantum mechanics, ii) the locality, iii) the freedom of choice on the measurement settings and iv) the observer-independent experimental outcomes.
In the present work, assumption i) has been accepted from the beginning considering all systems described by quantum mechanics (including detectors and human beings). 
Locality also has been implicitly assumed by considering only unitary processes (no projections or collapses that could induce ``spooky actions at a distance'') and by the form of the Hamiltonian considered here.
The freedom of choice is also implicitly assumed in the basic properties of the probability function required from the Gleason-Busch theorem.
From these considerations, it is no surprise that the
above probability expressions indicate a clear violation of the measurement output universality.
If we assume that the experimental outputs are relative to each pair of observer and observed system, we can have a different description of a same system from different observers, with no real paradox in the Wigner's friend scenario.


\section{Conclusions} \label{sec:conclusions}

We presented here a derivation of the conditional probability $\wp$ for subsequent measurements in the context of the quantum time approach.
This is obtained with a minimal number of assumptions, considering only systems subject to the rules of Quantum Mechanics, without classical systems, and interacting with each other by unitary operators with the implementation of POVM.
Applying the Gleason-Busch theorem to the global kinetic Hilbert space, $\wp$ is built from first principles.
With the approach 
 presented here with a one clock system, it emerges that a two-time probability function cannot be defined.
The causal order of the subsequent measurements is encoded in the Hamiltonian, whose structure should be considered as prior knowledge in the probability function.

When the Wigner's friend scenario is considered, the relativity of measurements naturally emerges, without contradictions or ambiguities. 
The roles of Wigner and his friend are completely interchangeable, and Wigner is seen by the friend as being 
in a superposition.
This is particularly evident in the marginal probabilities $\wp(w  \land t)$ and $\wp(f  \land t)$ and, as consequence, in the conditional probabilities $\wp(w | f \land t)$ and $\wp(f | w \land t)$.

Future developments will be focused on the extension of the formalism presented here for other Wigner's friend-type scenarios, like those considered for the no-go theorems in Refs~\cite{Frauchiger2018,Brukner2018,Bong2020}. 

\begin{acknowledgements}
I would like to thank N. Paul, M. Romanelli and 
E.~Lamour
for their support. 
I also wish to express my gratitude to the organizers of the QISS group who, by making available on-line their workshops and seminars  presentations, gave me the initial stimulus for this present work.
I would like to thank also ``referee n.2'' for the very valuable suggestions that considerably improve the original manuscript.
\end{acknowledgements}

\bibliography{qt}

\end{document}